\documentclass[12pt]{article}
\usepackage{epsfig}

\begin{document}

\begin{center}

\vspace*{1.0cm}

{\Large \bf{Search for double $\beta$ decay processes in
$^{106}$Cd with the help of $^{106}$CdWO$_4$ crystal
scintillator}}

\vskip 1.0cm

{\bf P.~Belli$^{a}$, R.~Bernabei$^{a,b,}$\footnote{Corresponding
author. {\it E-mail address:} rita.bernabei@roma2.infn.it
(R.~Bernabei).}, R.S.~Boiko$^{c}$, V.B.~Brudanin$^{d}$,
F.~Cappella$^{e,f}$, V.~Caracciolo$^{g,h}$, R.~Cerulli$^{g}$,
D.M.~Chernyak$^{c}$, F.A.~Danevich$^{c}$, S.~d'Angelo$^{a,b}$,
E.N.~Galashov$^{i}$, A.~Incicchitti$^{e,f}$, V.V.~Kobychev$^{c}$,
M.~Laubenstein$^{g}$, V.M.~Mokina$^{c}$, D.V.~Poda$^{g,c}$,
R.B.~Podviyanuk$^{c}$, O.G.~Polischuk$^{c}$, V.N.~Shlegel$^{i}$,
Yu.G.~Stenin$^{i,}$\footnote{Deceased}, J.~Suhonen$^{j}$, V.I.~Tretyak$^{c}$,
Ya.V.~Vasiliev$^{i}$}
 \vskip 0.3cm

 $^{a}${\it INFN, Sezione di Roma ``Tor Vergata'', I-00133 Rome, Italy}

 $^{b}${\it Dipartimento di Fisica, Universit\`a di Roma ``Tor Vergata'', I-00133 Rome, Italy}

 $^{c}${\it Institute for Nuclear Research, MSP 03680 Kyiv, Ukraine}

 $^{d}${\it Joint Institute for Nuclear Research, 141980 Dubna, Russia}

 $^{e}${\it INFN, Sezione di Roma ``La Sapienza'', I-00185 Rome, Italy}

 $^{f}${\it Dipartimento di Fisica, Universit\`a di Roma ``La Sapienza'', I-00185 Rome, Italy}

 $^{g}${\it INFN, Laboratori Nazionali del Gran Sasso, I-67100 Assergi (AQ), Italy}

 $^{h}${\it Dipartimento di Fisica, Universit$\grave{a}$ dell'Aquila, I-67100 L'Aquila, Italy}

 $^{i}${\it Nikolaev Institute of Inorganic Chemistry, 630090 Novosibirsk, Russia}

 $^{j}${\it Department of Physics, University of Jyv$\ddot{a}$skyl$\ddot{a}$, P.O. Box 35 (YFL), FI-40014 Finland}

\end{center}

\vskip 0.5cm

\begin{abstract}

A search for the double $\beta$ processes in $^{106}$Cd was carried out
at the Gran Sasso National Laboratories of the INFN (Italy)
with the help of a $^{106}$CdWO$_4$ crystal scintillator (215 g)
enriched in $^{106}$Cd up to 66\%. After 6590 h of data
taking, new improved half-life limits on the double beta decay processes in
$^{106}$Cd were established at the level of $10^{19}-10^{21}$ yr;
in particular, $T_{1/2}^{2\nu\varepsilon\beta^+}\geq 2.1\times
10^{20}$ yr, $T_{1/2}^{2\nu2\beta^+}\geq 4.3\times 10^{20}$ yr,
and $T_{1/2}^{0\nu2\varepsilon}\geq 1.0\times 10^{21}$ yr.
The resonant neutrinoless double electron captures to the 2718 keV, 2741 keV and 2748 keV
excited states of $^{106}$Pd are restricted to $T_{1/2}^{0\nu2K}\geq4.3\times 10^{20}$~yr,
$T_{1/2}^{0\nu KL_{1}}\geq9.5\times 10^{20}$~yr and $T_{1/2}^{0\nu KL_{3}}\geq4.3\times 10^{20}$~yr,
respectively (all limits at 90\% C.L.).
A possible resonant enhancement of the $0\nu2\varepsilon$
processes is estimated in the framework of the QRPA approach.
The radioactive contamination of the $^{106}$CdWO$_4$ crystal
scintillator is reported.

\end{abstract}

\vskip 0.4cm


\vskip 0.4cm


\section{INTRODUCTION}

The neutrinoless double beta decay ($0\nu2\beta$) is a powerful tool
to investigate the properties of the neutrino and of the weak interactions.
The study of this nuclear decay, forbidden in the framework of
the Standard Model, can allow us to determine an absolute scale of the
neutrino mass and its hierarchy, to establish the nature of the neutrino
(Majorana or Dirac particle), and to check the lepton number
conservation, the possible contribution of right-handed admixture to the
weak interaction, and the existence of Nambu-Goldstone bosons (majorons)
\cite{2bRev}.

Experimental efforts over the last seventy years have concentrated
mainly on the decay modes with emission of two electrons. Allowed in
the Standard Model, the two neutrino (2$\nu$) 2$\beta^-$ decay mode was
observed in ten isotopes with half-lives in the range of
$10^{18}-10^{24}$ yr. For the $0\nu2\beta^-$ decay mode
half-life limits at the level of
$10^{23}-10^{25}$ yr were set for
several nuclei (see reviews \cite{DBD-tab,Bara10} and original
studies \cite{HM,IGEX,Bern02,Dane03b,Cuoricino,NEMO-3}), while
positive evidence for $^{76}$Ge has been published in \cite{kk} and
new experiments are in progress to further investigate this latter isotope as well.

The results of the searches for the capture of two electrons from atomic
shells ($2\varepsilon$), electron capture with positron emission
($\varepsilon\beta^+$), and emission of two positrons ($2\beta^+$) are
at the level of $10^{16}-10^{21}$ yr (see review \cite{DBD-tab}
and original works
\cite{Bara07a,Kim07,Bara08,Daws08,Bell08a,Bell08b,Bara09,Bell09a,Bell09b,Bell09c,Gavr11,Rukh11a,Rukh11b,Bell11a,Andr11,Bara11,Bell11b});
although allowed, the two neutrino mode of these processes has not yet been
detected\footnote{An indication for $2\beta ^{+}$ decay processes
in $^{130}$Ba was obtained by the geochemical method
\cite{Mesh01,Pujo09}; however, this result has to be confirmed in
a direct counting experiment. It is worth mentioning the work
\cite{Ceru04} where BaF$_2$ crystal scintillators were used to
search for double $\beta$ processes in $^{130}$Ba.}.
High sensitivity experiments to search for neutrinoless
$2\varepsilon$ and $\varepsilon \beta^+$ decays are also important because they
could clarify a contribution of right-handed admixtures in weak interactions
\cite{Hirs94}.

\begin{figure}[htb]
\begin{center}
\mbox{\epsfig{figure=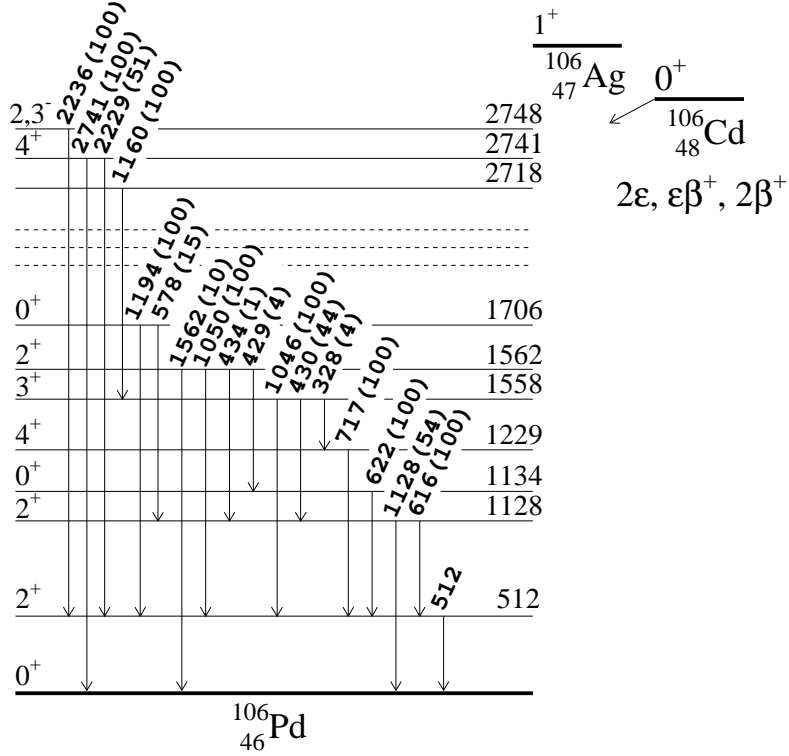,height=10.0cm}} \caption{Simplified
decay scheme of $^{106}$Cd \cite{Fren08} (levels at $1904-2714$
keV are omitted). The energies of the excited levels and of the emitted $\gamma$
quanta are in keV (relative intensities of $\gamma$ quanta are
given in parentheses).}
 \label{fig:106-Cd-scheme}
\end{center}
\end{figure}

The isotope $^{106}$Cd (the decay scheme is presented in Fig.
\ref{fig:106-Cd-scheme}) is among the most widely studied
$2\beta^+$ nuclides thanks to the large energy release ($Q_{2\beta}$ =
2775.39(10) keV \cite{Gon2011}) and to the comparatively high natural
abundance (1.25 $\pm$ 0.06\% \cite{Berg11}). It should be stressed
that $^{106}$Cd is a rather promising isotope also according to
the theoretical predictions
\cite{Hirs94,Stau91,Toiv97,Stoi03,Shuk05,Domi05}. In particular,
the calculated half-lives for the two neutrino mode of
the 2$\varepsilon$ and $\varepsilon\beta^{+}$ processes are at the
level of $T_{1/2}\sim 10^{20}-10^{22}$ yr
\cite{Toiv97,Bara96a,Rumy98,Civi98,Suho01}, reachable with the
present low counting technique.

Furthermore, in the case of the 0$\nu$ capture of two electrons from the
$K$ shell (or $L$ and $K$ shells), the energy releases of 2727
keV ($2K$ capture), 2747 keV ($KL_1$) and 2748 keV ($KL_3$) are close to the energies of a few
excited levels of $^{106}$Pd (with $E_{exc}=2718$ keV, 2741 keV and 2748 keV).
Such a coincidence could give a resonant enhancement of the
0$\nu2\varepsilon$ capture
\cite{Wint55b,Volo82,Bern83,Sujk02,Sujk04,Kriv11}.

Therefore, it is not surprising that the study of $^{106}$Cd has a rather long
history. The half-life limits at the level of
$10^{15}$ yr could be extracted from the old (1952) underground
measurements of a Cd sample with photographic emulsions
\cite{Frem52}, while a search for positrons emitted in $2\beta^+$
decay was performed in 1955 with a Wilson cloud chamber in
a magnetic field and with 30 g of cadmium foil; this gave a limit of
$10^{16}$ yr \cite{Wint55a}. Measurements of a 153 g Cd sample
during 72 h with two NaI(Tl) scintillators working in coincidence
have been carried out in \cite{Norm84}; the half-life limits at the
level of $\sim10^{17}$ yr were determined for $2\beta^+$,
$\varepsilon\beta^+$ and $2\varepsilon$ processes.

The subsequent studies can be divided into two groups:
experiments using samples of cadmium with external detectors for
the detection of the emitted particles (with enriched
$^{106}$Cd \cite{Rukh11a,Bell99} and natural cadmium
\cite{Bara96a}), and experiments with detectors containing
cadmium, namely semiconductor CdTe and CdZnTe detectors
\cite{Ito97,Daws09} and CdWO$_4$ crystal scintillators
\cite{Dane03b,Geor95,Dane96a}.
Previous experiments on the searches for the $2\beta$ processes in
$^{106}$Cd are summarized in Table \ref{tab:01}.

Data from the experiment, performed in the Solotvina Underground Laboratory (1000 m w.e.), 
with a 15 cm$^{3}$ $^{116}$CdWO$_4$
crystal scintillator (enriched in $^{116}$Cd to 83\%, with 0.16\%
of $^{106}$Cd), 
were used to set the limits on the 2$\beta$ decay
of $^{106}$Cd at the level of $10^{17}-10^{19}$ yr \cite{Geor95}.
In experiment \cite{Bara96a}, 331 g of Cd foil were measured
at the Frejus Underground Laboratory (4800 m w.e.) with a 120
cm$^3$ HPGe detector during 1137 h; $\gamma$ quanta from the
annihilations of the positrons and from the de-excitation of the
daughter $^{106}$Pd nucleus were searched for, giving rise to
half-life limits at the level of $10^{18}-10^{19}$ yr. In
\cite{Dane96a}, a large (1.046 kg) CdWO$_4$ scintillator was
measured at the Gran Sasso National Laboratories (3600 m w.e.)
over 6701 h. The determined limits on the half-life for the
$2\beta^+$ and $\varepsilon\beta^+$ decays were at the level of
$\sim10^{19}$ yr for $0\nu$, and $\sim 10^{17}$ yr for $2\nu$
processes. A small (0.5 g) CdTe crystal was tested as a cryogenic
bolometer in 1997 \cite{Ito97}; the achieved sensitivity was
$\sim10^{16}$ yr for $0\nu2\beta^+$ decay. An experiment
\cite{Bell99} was performed in 1999 at the Gran Sasso National
Laboratories using an enriched $^{106}$Cd (to 68\%) cadmium sample
(154 g) and two low background NaI(Tl) scintillators installed in
the low background DAMA/R\&D set-up during 4321 h; these
measurements reached a sensitivity level of more than $10^{20}$ yr
for $2\beta^+$, $\varepsilon\beta^{+}$ and $2\varepsilon$
processes. A long-term (14183 h) experiment in the Solotvina
Underground Laboratory with enriched $^{116}$CdWO$_4$
scintillators (total mass of 330 g) was completed in 2003
\cite{Dane03b}; in addition, results of dedicated measurements
during 433 h with a 454 g not-enriched CdWO$_4$ crystal were also
considered \cite{Dane96b}. In general, the experimental
sensitivity was improved by approximately one order of magnitude
in comparison with the older measurements \cite{Geor95}.

There are two running experiments to search for 2$\beta$ decay of
$^{106}$Cd: COBRA and TGV-II. The $T_{1/2}$ limits in the range of
$10^{17}-10^{18}$ yr were set in the COBRA experiment
\cite{Daws09} using CdTe and CdZnTe crystals. In the TGV-II
experiment \cite{Rukh11a,Rukh11b}, 32 planar HPGe detectors are
used. Cadmium foils enriched in $^{106}$Cd to 75\% are inserted
between neighbouring detectors. The main goal of the TGV
experiment is the search for the two neutrino double electron capture in
$^{106}$Cd. After 8687 h plus 12900 h (in two phases of the
experiment) of data taking, the limits on double $\beta$ decay of
$^{106}$Cd to the ground state and to the excited levels of $^{106}$Pd
are around $10^{20}$ yr.

\nopagebreak
\begin{table*}[htb]
\caption{Experiments on the searches for the $2\beta$ decay of $^{106}$Cd.
The range of the $T_{1/2}$ limits corresponds to values given for the
transitions to the ground state or to the excited levels of
$^{106}$Pd. More detailed information can be found in the original
papers (see also \cite{DBD-tab}). COBRA and TGV experiments are
still running.}
\begin{center}
\resizebox{0.93\textwidth}{!}{
\begin{tabular}{|l|l|l|}

 \hline
 {\bf Description}                                          & {\bf $T_{1/2}$ limit, yr}                       & {\bf Year [Ref.]}\\
 \hline
 Cd samples between photographic emulsions$^{a}$           & $\sim10^{15}$ ($0\nu2\beta^{+}$, $0\nu\varepsilon\beta^{+}$) & 1952 \cite{Frem52}\\

 \hline
 Cd foil in a Wilson cloud chamber                          & $6\times10^{16}$ ($0\nu2\beta^+$)         & 1955 \cite{Wint55a}\\

 \hline
 Cd sample between two NaI(Tl) scintillators in coincidence & ($2.2-2.6)\times10^{17}$ ($2\beta^+$)                     & 1984 \cite{Norm84}\\
 ~                                                          & ($4.9-5.7)\times10^{17}$ ($\varepsilon\beta^{+}$)         & \\
 ~                                                          & $1.5\times10^{17}$ ($2\nu2\varepsilon$)                   & \\

 \hline
 $^{116}$CdWO$_4$ crystal scintillator                      & ($0.5-1.4)\times10^{18}$ ($0\nu2\beta^{+}$)               & 1995 \cite{Geor95} \\
 ~                                                          &  ($0.3-1.1)\times10^{19}$ ($0\nu\varepsilon\beta^{+}$)    & ~ \\
 ~                                                          &  5.8$\times$10$^{17}$ ($2\nu2\varepsilon$)                & ~\\
 \hline
 CdWO$_4$ crystal scintillator                              & $2.2\times10^{19}$ ($0\nu2\beta^{+}$)                     & 1996 \cite{Dane96a}\\
 ~                                                          & $9.2\times10^{17}$ ($2\nu2\beta^{+}$)                     & ~ \\
 ~                                                          & $5.5\times10^{19}$ ($0\nu\varepsilon\beta^+$)             & ~ \\
 ~                                                          & $2.6\times10^{17}$ ($2\nu\varepsilon\beta^+$)             & ~ \\
 \hline
 Cd sample measured by HPGe detector                                 & $1.0\times10^{19}$ ($2\beta^+$)                  & 1996 \cite{Bara96a} \\
 ~                                                          & ($6.6-8.1)\times10^{18}$ ($\varepsilon\beta^+$)           & ~ \\
 ~                                                          & ($3.5-6.2)\times10^{18}$ ($2\varepsilon$)                 & ~ \\
 \hline
  CdTe cryogenic bolometer                                  & $1.4\times10^{16}$ ($0\nu\varepsilon\beta^+$)             & 1997 \cite{Ito97} \\
 \hline
 $^{106}$Cd sample between two NaI(Tl) scintillators in coincidence & (1.6$-$2.4)$\times$10$^{20}$ ($2\beta^+$)         & 1999 \cite{Bell99}\\
 ~                                                          & ($1.1-4.1)\times10^{20}$ ($\varepsilon\beta^+$)           & ~ \\
 ~                                                          &  ($3.0-7.3)\times 10^{19}$ ($2\varepsilon)$               & ~ \\
 \hline
 $^{116}$CdWO$_4$ crystal scintillators                     & ($0.5-1.4)\times10^{19}$ ($2\beta^+$)                     & 2003 \cite{Dane03b} \\
 ~                                                          & ($0.1-7.0)\times10^{19}$ ($\varepsilon\beta^+$)           & ~ \\
 ~                                                          &  ($0.6-8.0)\times10^{18}$ ($2\varepsilon$)                & ~ \\
 \hline
 CdZnTe semiconductor detectors (COBRA)                     & ($0.9-2.7)\times10^{18}$ ($2\beta^+$)                     & 2009 \cite{Daws09} \\
 ~                                                          & ($4.6-4.7)\times10^{18}$ ($\varepsilon\beta^+$)           & ~ \\
 ~                                                          &  $1.6\times10^{17}$ ($2\varepsilon$)                      & ~ \\
 \hline
  $^{106}$Cd samples between planar HPGe detectors (TGV)    & $3.6\times10^{20}$ ($2\nu2\varepsilon$)                   & 2011 \cite{Rukh11a} \\
 ~                                                          & $1.1\times10^{20}$ ($0\nu2\varepsilon$, 2741 keV)         & ~ \\
 \cline{2-3}
 ~                                                          & ($1.4-1.7)\times10^{20}$ ($2\beta^+$)                     & 2011 \cite{Rukh11b}\\
 ~                                                          & ($1.1-1.6)\times10^{20}$ ($\varepsilon\beta^{+}$)         & ~ \\
 ~                                                          & $1.6\times10^{20}$ ($0\nu2\varepsilon$, 2718 keV)         & ~ \\
 \hline
 \multicolumn{3}{l}{$^{a)}$ To our knowledge, this was the first underground experiment in history of investigations of 2$\beta$ decay.} \\
 \end{tabular}
 }
 \label{tab:01}
 \end{center}
 \end{table*}

We would like to mention two important advantages of the experiments
using detectors containing cadmium: a higher detection efficiency
for the different channels of the $^{106}$Cd double $\beta$ decay, and
a possibility to resolve the two neutrino and the neutrinoless modes
of the decay.

Thanks to their good scintillation characteristics, their low level of intrinsic
radioactivity, and their pulse-shape discrimination ability (which allows an
effective reduction of the background),
cadmium tungstate crystal scintillators were successfully applied
to low background experiments in order to search for the double $\beta$ decay
of the cadmium and tungsten isotopes \cite{Dane03b,Bell08b,Dane96a},
and in order to investigate rare $\alpha$ \cite{Dane03a} and $\beta$
\cite{Dane96b,Bell07a} decays.

The aim of the present work was the search ~for the $2\beta$ processes in
$^{106}$Cd with the help of a low background cadmium tungstate
crystal scintillator enriched in $^{106}$Cd
($^{106}$CdWO$_4$).

\section{EXPERIMENT}
\label{exp}

The cadmium tungstate crystal (27 mm in diameter by 50 mm in length;
mass 215 g), used in the experiment, was developed \cite{Bell10a} from deeply purified cadmium
\cite{Kovt11} enriched in $^{106}$Cd to 66\%.
The scintillator was fixed inside a cavity
($\oslash47\times59$ mm) in the central part of a polystyrene
light-guide, 66 mm in diameter by 312 mm in length. The cavity was
filled with high purity silicon oil. Two high purity quartz
light-guides, 66 mm in diameter by 100 mm in length, were optically
connected to the opposite sides of the polystyrene light-guide. To
collect the scintillation light the assembly was viewed by two low
radioactive EMI9265--B53/FL, 3'' diameter photomultiplier tubes (PMT).
The detector was installed deep underground in the low background
DAMA/R\&D set-up at the Gran Sasso National Laboratories of the
INFN (Italy). It was surrounded by copper bricks and sealed in a low
radioactive, air-tight copper box continuously flushed with high purity
nitrogen gas to avoid the presence of residual environmental radon.
The copper box was surrounded by a passive shield made of high purity
copper, 10 cm of thickness, 15 cm of low radioactive lead, 1.5 mm of
cadmium and 4 to 10 cm of polyethylene/paraffin to reduce the
external background. The shield was contained inside a Plexiglas
box, also continuously flushed with high purity nitrogen gas.

An event-by-event data acquisition system recorded the amplitude,
the arrival time, and the pulse shape of the events by means of a 1 GS/s 8 bit DC270
Transient Digitizer by Acqiris (adjusted to a sampling frequency
of 20 MS/s) over a time window of 100 $\mu$s.

The energy resolution of the detector was measured with $^{22}$Na, $^{60}$Co,
$^{133}$Ba, $^{137}$Cs, $^{228}$Th, and $^{241}$Am $\gamma$
sources in the beginning of the experiment. For instance, the
energy resolution (full width at half maximum, FWHM) of the
$^{106}$CdWO$_4$ detector for the $\gamma$ quanta of
$^{137}$Cs (662 keV) and of $^{228}$Th (2615 keV) was 14.2(3)\% and 8.4(2)\%,
respectively. Two additional calibration measurements were
performed: one approximately in the middle, and the second one at
the end of the experiment with the help of $^{22}$Na, $^{60}$Co, $^{137}$Cs,
and $^{228}$Th $\gamma$ sources to test the detector stability. In
addition, the energy scale of the detector was checked by using
the peaks due to $^{207}$Bi contamination of the $^{106}$CdWO$_4$
crystal scintillator (see Section \ref{bg-sim}). The
energy scale during the experiment was reasonably stable with
a deviation in the range of $(1-2)\%$. The data of the calibration
measurements were used to estimate the dependence of the energy
resolution on the energy. Below 500 keV the energy resolution of the
detector to $\gamma$ quanta with energy $E_{\gamma}$ can be
described by the function: FWHM$_\gamma$ = $\sqrt{11.2\times
E_{\gamma}}$, while above 500 keV the data are fitted by
FWHM$_{\gamma}$ = $\sqrt{-4900+21\times E_{\gamma}}$, where
FWHM$_\gamma$ and $E_{\gamma}$ are given in keV.

The low background measurements were carried out in three runs
listed in Table \ref{runs}. The energy interval of the data taking was
chosen as $0.05-4$ MeV in Run 1 to investigate the background of
the detector at low energy. Taking into account the rather high
activity of $\beta$ active $^{113}$Cd$^m$  (see the next Section), the
data acquisition was slightly modified in order to avoid the recording of
the pulse shapes of all events with an energy lower than 0.4 MeV
(Run 2); the upper energy threshold was $\approx 1.8$ MeV.
In a third run, after some improvement in the data acquisition system,
the energy threshold was increased
to $\approx$ 0.57 MeV and the upper energy threshold was set to 4 MeV
(Run 3). The data accumulated in Run 2 were used to estimate
the activity of $^{228}$Th in the $^{106}$CdWO$_4$ crystal by a
time-amplitude analysis (see Section \ref{t-A}). The first 1320 h
of data taking (Run 1 + part of Run 3) were already analyzed
and presented in \cite{Bell10b}.

\begin{table}[htb]
\caption{The low background measurements with the $^{106}$CdWO$_4$
crystal scintillator. Times of measurements ($t$), energy
intervals of data taking ($\Delta E$), and background counting
rates (BG) in different energy intervals are specified.}
\begin{center}
\begin{tabular}{|c|c|l|c|c|c|}
\hline
  Run   & $t$   & $\Delta E$    & \multicolumn{3}{|c|}{BG (counts/(yr$\times$keV$\times$kg))} \\
  ~     & (h)   & (MeV)         & \multicolumn{3}{|c|}{in energy interval (MeV)} \\
 \cline{4-6}
  ~     & ~     & ~             & $0.8-1.0$     & $2.0-2.9$     & $3.0-4.0$ \\
 \hline
  1     & 283   & $0.05-4.0$    & 474(18)       & 2.6(6)        & 0.4(3) \\
  2     & 2864  & $0.40-1.8$    & 453(11)       & --            & -- \\
  3     & 6307  & $0.57-4.0$    & 412(4)        & 2.3(1)        & 0.33(4) \\
 \hline
\end{tabular}
\label{runs}
\end{center}
\end{table}

\section{DATA ANALYSIS}

\begin{figure}[htb]
\begin{center}
\mbox{\epsfig{figure=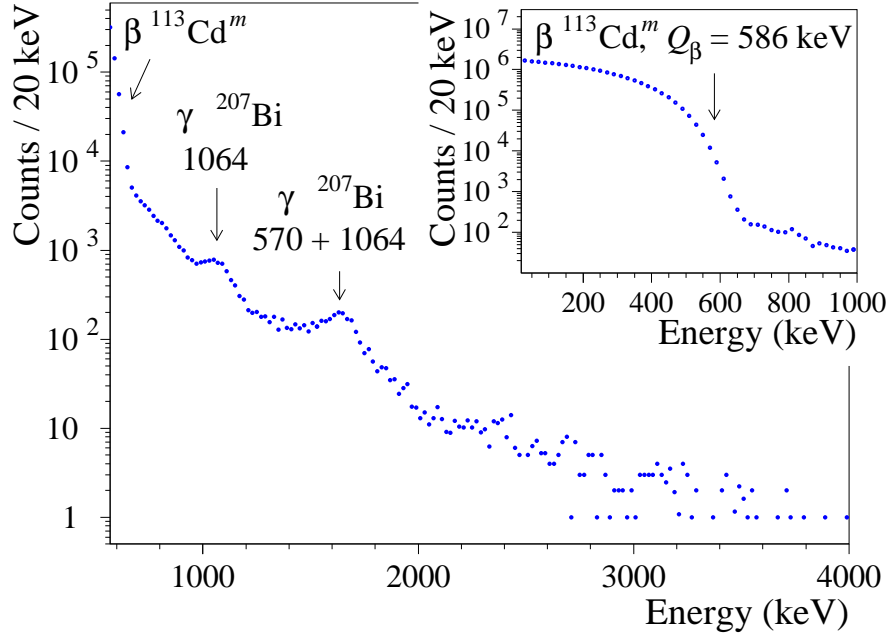,height=9.0cm}} \caption{(Color online) The energy
spectrum measured with the $^{106}$CdWO$_4$ scintillator over 6590
h in the low background set-up. (Inset) The decay of the $\beta$ active
$^{113}$Cd$^m$ dominates at the energy $<0.65$ MeV (the data over 283
h).}
 \label{fig:BG-raw}
\end{center}
\end{figure}

The energy spectrum accumulated with the $^{106}$CdWO$_4$ detector
in Runs 1 and 3 over 6590 h is presented in Fig.
\ref{fig:BG-raw}. The counting rate $\approx 24$ counts/s below
the energy $\approx0.65$ MeV is mainly due to the $\beta$ decay
of $^{113}$Cd$^m$ with activity 116(4) Bq/kg. Contamination of
the enriched $^{106}$Cd by the $\beta$ active $^{113}$Cd$^m$ has been
found in the low background TGV experiment \cite{Rukh06}, where
$\beta$ particles and X rays from thin foils of the enriched
$^{106}$Cd were measured by planar Ge detectors; part of this
material was used to produce the $^{106}$CdWO$_4$ crystal.

Contributions to the background above the energy $\approx 0.6$ MeV were
analyzed by means of the time-amplitude and of the pulse-shape
discrimination techniques, as well as by the fit of the data with Monte
Carlo simulated models of the background.

\subsection{Time-amplitude analysis of $^{228}$Th activity}
\label{t-A}

\begin{figure}[htb]
\begin{center}
\mbox{\epsfig{figure=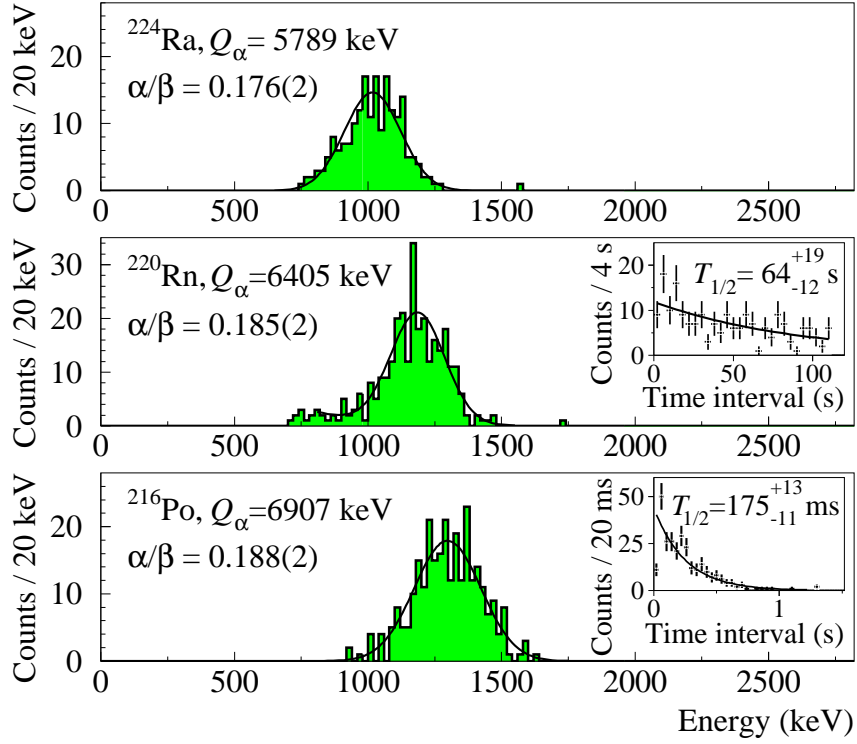,height=10.0cm}} \caption{(Color
online) Alpha peaks of $^{224}$Ra, $^{220}$Rn and $^{216}$Po
selected by the time-amplitude analysis from the data accumulated
over 9454 h with the $^{106}$CdWO$_4$ detector. The obtained
half-lives of $^{220}$Rn ($64^{+19}_{-12}$ s) and $^{216}$Po
($175^{+13}_{-11}$ ms) are in agreement with the table
values (55.6 s and 145 ms, respectively \cite{ToI98}).}
\label{fig:t-A}
\end{center}
\end{figure}

The arrival time and the energy of each event were used to select the
events of the fast decay chain in the $^{232}$Th
family\footnote{The technique of the time-amplitude analysis is
described in detail e.g. in \cite{Dane95,Dane01}.}: $^{224}$Ra
($Q_\alpha $ = $5.79$~MeV, $T_{1/2}$ = $3.66$~d) $\rightarrow$
$^{220}$Rn ($Q_\alpha $ = $6.41$~MeV, $T_{1/2}$ = $55.6$~s)
$\rightarrow$ $^{216}$Po ($Q_\alpha $ = $6.91$ MeV, $T_{1/2}$ =
$0.145$~s) $\rightarrow $ $^{212}$Pb. To select $\alpha$ events
from the decays of $^{224}$Ra, $^{220}$Rn, and $^{216}$Po, one
should take into account the quenching of the scintillation output in the
CdWO$_4$ crystal scintillator, the so called $\alpha/\beta$ ratio,
defined as the ratio of an $\alpha$ peak position in the $\gamma$
scale of a detector to the energy of the alpha particles. The
dependence of the $\alpha/\beta$ ratio on the energy of the $\alpha$
particles measured for $^{116}$CdWO$_4$ scintillator \cite{Dane03a}:
$\alpha/\beta=0.083(9)+0.0168(13)\times E_{\alpha}$
(where $E_{\alpha}$ is in MeV), was used to estimate the positions of
$^{224}$Ra, $^{220}$Rn, and $^{216}$Po $\alpha$ peaks in the data
accumulated with the $^{106}$CdWO$_4$ detector. As a first step,
all the events within an energy interval $0.6-1.8$ MeV were used as
triggers, while for the second events
a time interval $0.026-1.45$ s and the same energy
window were required. Taking into account the
efficiency of the events selection in this time interval (88.2\%
of $^{216}$Po decays), the activity of $^{228}$Th in the
$^{106}$CdWO$_4$ crystal was calculated to be 0.042(4)~mBq/kg. As a
next step, all the selected pairs ($^{220}$Rn~--~$^{216}$Po) were used
as triggers in order to find the events of the decay of the mother
$\alpha$ active $^{224}$Ra. A $1.45-111$ s time interval (73.2\%
of $^{220}$Rn decays) was chosen to select events in the energy
interval $0.6-1.75$ MeV. The obtained $\alpha$ peaks from the
$^{224}$Ra$\rightarrow^{220}$Rn$\rightarrow^{216}$Po$\rightarrow^{212}$Pb
chain and the time distributions for the
$^{220}$Rn$\rightarrow^{216}$Po and
$^{216}$Po$\rightarrow^{212}$Pb decays are shown in Fig.
\ref{fig:t-A}.

\nopagebreak

The positions of the three $\alpha$ peaks, selected by the
time-amplitude analysis in the $\gamma$ scale of the detector, were
used to obtain the following dependence of the $\alpha/\beta$ ratio on
the energy of the $\alpha$ particles, $E_{\alpha}$, in the range
$5.8-6.9$~MeV: $\alpha/\beta = 0.11(2)+0.011(3)\times E_{\alpha}$
(where $E_{\alpha}$ is in MeV). The dependence is in agreement
with the data obtained for the $^{116}$CdWO$_4$ scintillation detector
in \cite{Dane03a}.

\subsection{Pulse-shape discrimination}
\label{PSD}

\begin{figure}[htb]
\begin{center}
\mbox{\epsfig{figure=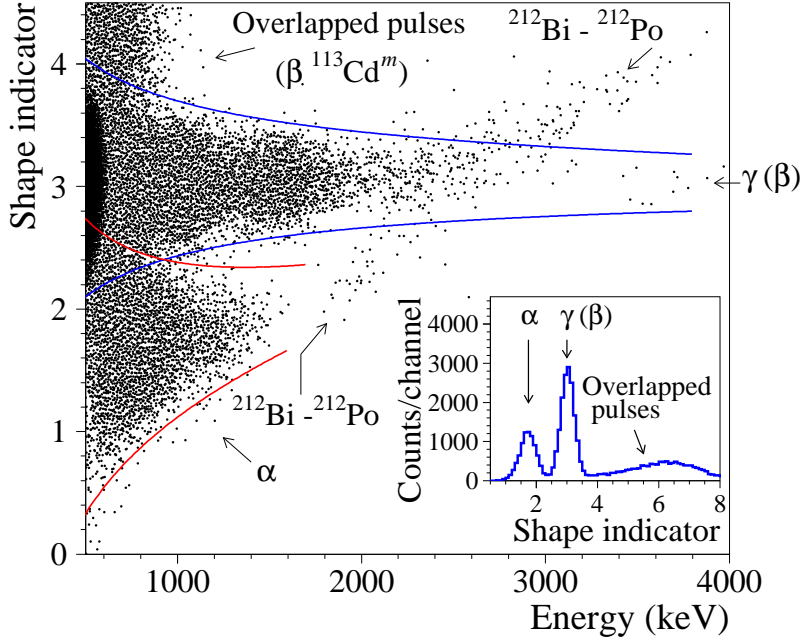,height=9.0cm}}
\caption{(Color
online) The shape indicators (see text) versus the energy accumulated over
6590 h with the $^{106}$CdWO$_4$ crystal scintillator in the low
background set-up. Three sigma intervals for shape indicator
values corresponding to $\gamma$ quanta ($\beta$ particles) and
$\alpha$ particles are depicted. Events with shape indicator
values greater than $\approx3.8$ can be explained by the overlap of
events (mainly of $\beta$ decays of $^{113}$Cd$^m$ in the crystal),
while the population of the events in the energy interval $\approx
1.8-3.8$ MeV with shape indicators outside of the
$\gamma$($\beta$) region are due to the decays of the fast
$^{212}$Bi~--~$^{212}$Po sub-chain of $^{228}$Th. (Inset)
The distribution of the shape indicators demonstrates the efficiency of
the pulse-shape discrimination between $\gamma$($\beta$), $\alpha$
and overlapped pulses.} \label{fig:SI-vs-E}
\end{center}
\end{figure}

As demonstrated in \cite{Fazz98}, the difference in pulse shapes
in the CdWO$_4$ scintillator can be used to discriminate
$\gamma(\beta)$ events from those induced by $\alpha$ particles.
The optimal filter method proposed by E.~Gatti and F.~De~Martini
in 1962 \cite{Gatt62} was applied for this purpose. For each
signal $f(t)$, the numerical characteristic of its shape (shape
indicator, $SI$) was defined as: $SI=\sum f(t_k)\times P(t_k)/\sum
f(t_k)$. There the sum is over the time channels $k,$ starting
from the origin of signal and averaging up to 50 $\mu$s, and $f(t_k)$
is the digitized amplitude (at the time $t_k$) of a given signal.
The weight function $P(t)$ was defined as: $P(t)=\{{f}_\alpha
(t)-{f}_\gamma (t)\}/\{f_\alpha (t)+f_\gamma (t)\}$, where
$f_\alpha (t)$ and $f_\gamma (t)$ are the reference pulse shapes
for $\alpha$ particles and $\gamma$ quanta measured in
\cite{Bard06}. By using this approach, $\alpha$ events were clearly
separated from $\gamma$($\beta$) events as shown in Fig.
\ref{fig:SI-vs-E} where the scatter plot of the shape indicator
versus energy is depicted for the data of the low background measurements with
the $^{106}$CdWO$_4$ detector. The distribution of the
shape indicators for events with energies in the range
$0.7-1.4$ MeV (shown in Inset of Fig. \ref{fig:SI-vs-E}) justifies
reasonable pulse-shape discrimination between $\alpha$ particles
and $\gamma$ quanta ($\beta$ particles), as well as a possibility
to reject randomly overlapped pulses (mainly caused by the $\beta$
decay of $^{113}$Cd$^m$).

\nopagebreak

\begin{figure*}[htb]
\begin{center}
\resizebox{0.55\textwidth}{!}{\includegraphics{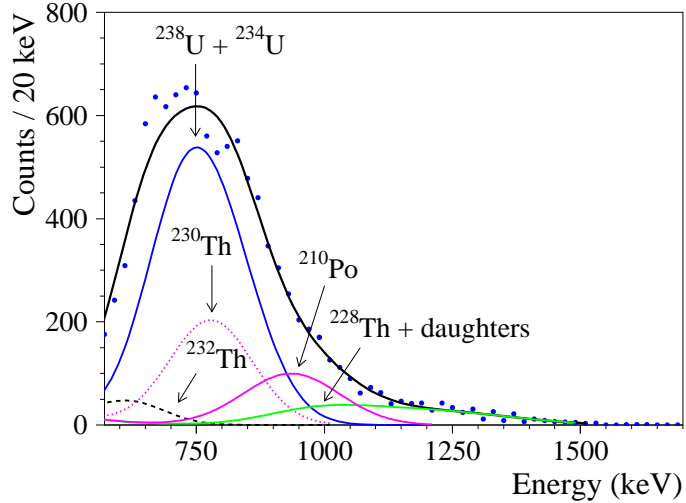}}
\caption{(Color online) The energy distribution of the $\alpha$ events
(points) selected by the pulse-shape analysis from the data
accumulated over 6590 h with the $^{106}$CdWO$_4$ detector
together with the fit (solid line) by a model, which includes the
$\alpha$ decays from $^{232}$Th and $^{238}$U families.}
 \label{fig:alpha}
\end{center}
\end{figure*}

The energy spectrum of the $\alpha$ events selected with the help of
the pulse-shape discrimination is shown in Fig. \ref{fig:alpha}.
As demonstrated in \cite{Dane03a}, the energy resolution
for the $\alpha$ particles is worse than that for the $\gamma$ quanta due
to the dependence of the $\alpha/\beta$ ratio on the direction of the
$\alpha$ particles relative to the CdWO$_4$ crystal
axes\footnote{One could compare the energy resolutions of the $\alpha$
peaks presented in Fig. \ref{fig:t-A} with the expected resolution
for $\gamma$ quanta (see Section \ref{exp}).}. As a result we
cannot definitively identify single U/Th $\alpha$ active daughters in the
spectrum. Therefore, we set only limits on $\alpha$ activities of
U/Th daughters in the $^{106}$CdWO$_4$ crystal scintillator. For
this purpose, the spectrum was fitted in the energy interval
$550-1500$ keV by a simple model, built of Gaussian functions
(to describe the $\alpha$ peaks of U/Th daughters) plus an exponential
function to describe the background. The activities of $^{228}$Th and
$^{226}$Ra were restricted taking into account the results of the
time-amplitude and of the double pulse (see Section \ref{bi-po})
analyses. The fit and its components are shown in Fig.
\ref{fig:alpha}. The limits on the activity of the U/Th daughters (supposing a
broken equilibrium in the chains) are presented in Table
\ref{rad-cont}. The total $\alpha$ activity of U/Th in the
$^{106}$CdWO$_4$ crystal is 2.1(2) mBq/kg.

The pulse-shape analysis also allows us to distinguish the main part of the
$^{212}$Bi$\rightarrow^{212}$Po$\rightarrow^{208}$Pb events from the
trace contamination of the crystal by $^{228}$Th (see Fig.
\ref{fig:SI-vs-E}).

\subsection{Identification of Bi-Po events}
\label{bi-po}

The search for the fast decays $^{214}$Bi ($Q_{\beta}=3.27$ MeV,
$T_{1/2}=19.9$ m) $\rightarrow$ $^{214}$Po ($Q_{\alpha}=7.83$ MeV,
$T_{1/2}=164~\mu$s) $\rightarrow$ $^{210}$Pb (in equilibrium with
$^{226}$Ra from the $^{238}$U chain) was performed with the help
of the pulse-shape analysis of the double pulses\footnote{The
technique of the analysis is described e.g. in
\cite{Dane03a,Bell07b}.}. Only eleven $^{214}$Bi~--~$^{214}$Po
events were found in the data over 6590 h. Taking into account the
detection efficiency in the time window of $1-50~\mu$s (it
contains 18.6\% of the $^{214}$Po decays), one can estimate the
activity of $^{226}$Ra in the $^{106}$CdWO$_4$ crystal as
0.012(3)~mBq/kg.

\begin{figure*}[htb]
\begin{center}
\resizebox{0.55\textwidth}{!}{\includegraphics{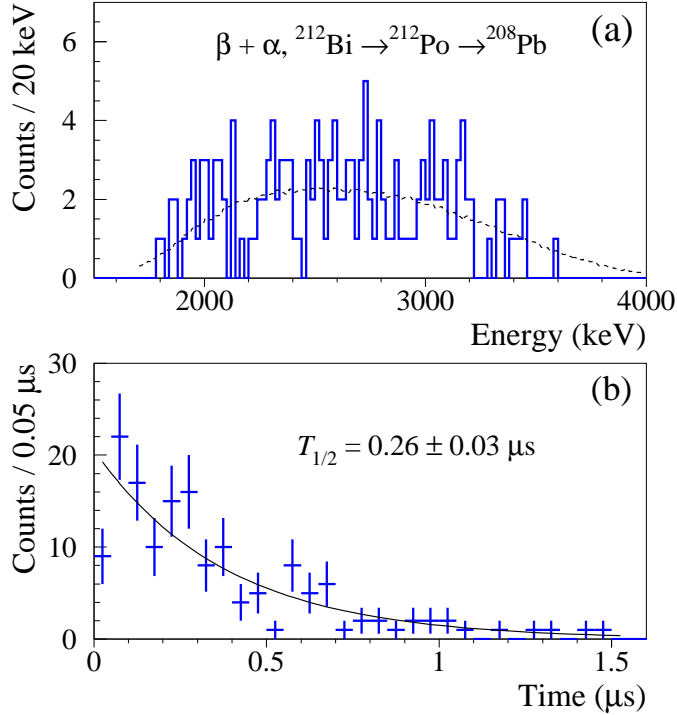}}
\caption{(Color online) (a) The energy spectrum of $^{212}$Bi$\to
^{212}$Po$\to ^{208}$Pb events in the $^{106}$CdWO$_4$
scintillator selected by means of the pulse-shape and of the front edge
analyses (see text) from the data accumulated over 6590 h together
with the fit (dashed line) of the simulated distribution. (b) The time
distribution of the $^{212}$Po $\alpha $ decay selected by the front
edge analysis. The fit of the time distribution gives a half-life:
$T_{1/2}=(0.26\pm 0.03)~\mu$s, in agreement with the table
value for $^{212}$Po (0.299 $\mu$s \cite{ToI98}). }
 \label{fig:Bi-Po}
\end{center}
\end{figure*}

To select double pulses produced by the fast chain of the decays
$^{212}$Bi ($Q_{\beta}=2.25$ MeV, $T_{1/2}=60.55$ m) $\rightarrow$
$^{212}$Po ($Q_{\alpha}=8.95$ MeV, $T_{1/2}=0.299~\mu$s)
$\rightarrow$ $^{208}$Pb (in equilibrium with $^{228}$Th from the
$^{232}$Th family), a front edge analysis was developed (see also
\cite{Dane03a}). The energy spectrum of the selected
$^{212}$Bi~--~$^{212}$Po events and the time distribution of
$^{212}$Po decay are presented in Fig. \ref{fig:Bi-Po}. The
approach gives the activity of $^{228}$Th as 0.051(4) mBq/kg, in a
reasonable agreement with the result of the time-amplitude
analysis.

All the selected Bi-Po events were removed from the $\gamma$($\beta$)
spectrum of the $^{106}$CdWO$_4$ detector.

\subsection{Simulation of the $\gamma(\beta)$ background, radioactive contamination of $^{106}$CdWO$_4$ scintillator}
\label{bg-sim}

To reproduce the background of the $^{106}$CdWO$_4$ detector, we
consider the contribution of the primordial radioactive isotopes $^{40}$K
and $^{238}$U/$^{232}$Th with their daughters, anthropogenic
radionuclides  $^{90}$Sr-$^{90}$Y and $^{137}$Cs, and cosmogenic
$^{106}$Ru and $^{110m}$Ag. Anthropogenic $^{90}$Sr and $^{137}$Cs are
the most widespread
radionuclides, in particular after the Chernobyl accident.
Contamination
of cadmium tungstate by $^{106}$Ru and $^{110m}$Ag was estimated in
\cite{bellini}, while
presence of $^{110m}$Ag in $^{116}$CdWO$_4$ crystal scintillators was
observed in \cite{Barab11}. The
radioactive contamination of the
set-up (in particular the PMTs and the copper box) can contribute
to the background, too. The energy distributions of the possible
background components were simulated with the help of the EGS4
\cite{EGS4} and GEANT4 \cite{GEANT} codes. The initial kinematics of the particles emitted
in the nuclear decays was given by the event generator DECAY0
\cite{DECAY4}.

The background energy spectrum of the $\gamma$ and $\beta$ events,
selected by means of the pulse-shape, of the front edge and of the double pulse
analyses, was fitted by a model built from the simulated
distributions. The activities of the U/Th daughters were bounded taking
into account the results of the time-amplitude and of the pulse-shape
analyses. The activities of the $^{40}$K, $^{232}$Th and $^{238}$U
in the PMTs were taken from \cite{Bern99}. The radioactive
contaminations of the copper box have been assumed to be equal to
those reported in \cite{Gun97}. In addition, we have added a model
of the overlapped $^{113}$Cd$^m$ $\beta$ decays, which contribute to the
background in the energy region up to $\approx 1$ MeV.

Two clear peculiarities in the spectrum of the CdWO$_4$
detector at $(1064\pm3)$ keV and at $(1631\pm5)$ keV cannot be explained
by the contribution from the external $\gamma$ quanta. Indeed, no
similar peaks were observed in the low background measurements
with radiopure ZnWO$_4$ crystal scintillators \cite{ZWO_bg}
performed before the present experiment in the same experimental
conditions. To explain the peculiarities, we suppose a pollution of
the crystal by $^{207}$Bi ($T_{1/2}=31.55$~yr, $Q_{EC}=2398$~keV
\cite{ToI98}). The presence of $^{207}$Bi could be caused by
the contamination of the facilities at the Nikolaev Institute of
Inorganic Chemistry (Novosibirsk, Russia) where the
$^{106}$CdWO$_4$ crystal was grown. A large amount of BGO crystal
scintillators is in production in that laboratory. BGO crystal
scintillators are typically contaminated by $^{207}$Bi at the
level of $0.01-10$ Bq/kg \cite{Baly93,Marc03,Coro08,Grig10}. Moreover, we
cannot also exclude the possibility of a $^{106}$CdWO$_4$ crystal
surface contamination in the laboratory of the Institute for
Nuclear Research (Kyiv, Ukraine) where the scintillator was
diffused and preliminary tested \cite{Bell10a} with several gamma
sources, including an open $^{207}$Bi source. Therefore, two
distributions of $^{207}$Bi (uniformly distributed in the crystal
volume and deposited on its surface) were also simulated and added
to the background model.

\begin{figure}[htb]
\begin{center}
\mbox{\epsfig{figure=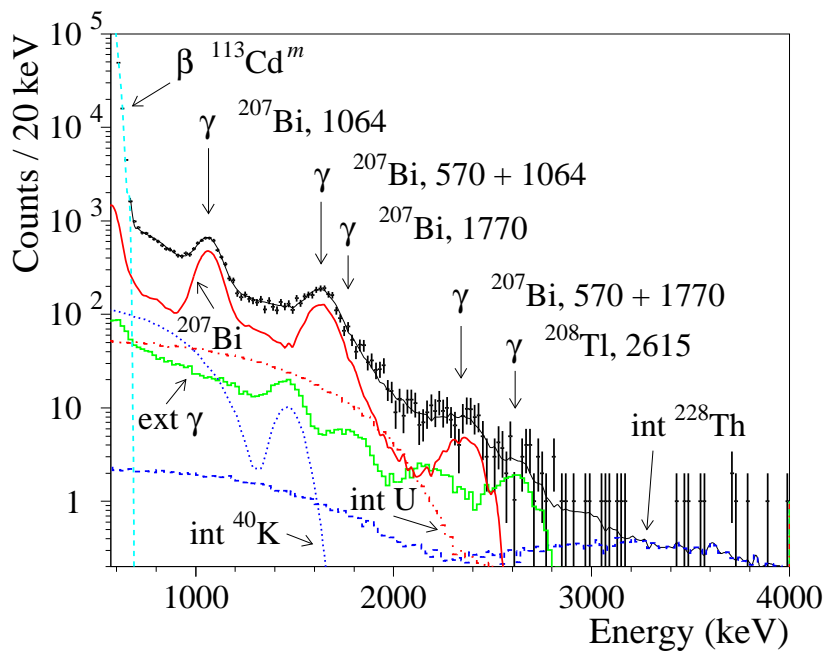,height=9.0cm}} \caption{(Color
online) The energy spectrum of the $\beta$($\gamma$) events
accumulated over 6590 h in the low background set-up with the
$^{106}$CdWO$_4$ crystal scintillator (points) together with the background model (black continuous superimposed line).
The main components of the background are shown:
the $\beta$ spectrum of the internal $^{113}$Cd$^m$, the distributions of
$^{40}$K, $^{228}$Th, $^{238}$U, $^{207}$Bi (deposited on the
crystal surface), and the contribution from the external $\gamma$
quanta from PMTs and copper box (``ext $\gamma$'') in these
experimental conditions.}
 \label{fig:BG-fit}
\end{center}
\end{figure}

A fit of the spectrum of $\gamma$($\beta$) events in the energy
region $0.66-4.0$ MeV by the model described above, and by the main
components of the background are shown in Fig. \ref{fig:BG-fit}.
The fit ($\chi^2/$n.d.f.~$=111/108=1.03$, where n.d.f. is number
of degrees of freedom) confirmed more likely a surface
contamination of the crystal scintillator by $^{207}$Bi at
level of 3 mBq (0.06 mBq/cm$^2$). We cannot distinguish the
part of the activity due to bulk contamination and we give only a
limit on the internal contamination of the crystal by $^{207}$Bi as
$\leq 0.7$ mBq/kg.

There are no other clear peculiarities in the spectrum which could
be ascribed to the internal trace radioactive contamination.
Therefore, we just set limits on the activities of $^{40}$K,
$^{90}$Sr-$^{90}$Y, cosmogenic $^{106}$Ru and $^{110m}$Ag. A
summary of radioactive contamination of the $^{106}$CdWO$_4$
crystal scintillator is given in Table \ref{rad-cont}. We hope to
clarify further the radioactive contamination of the scintillator
at a next stage of the experiment by running the $^{106}$CdWO$_4$
crystal scintillator in coincidence/anti-coincidence with an
ultra-low background HPGe $\gamma$ detector.

\nopagebreak
\begin{table}[htb]
\caption{Radioactive contamination of the $^{106}$CdWO$_4$
scintillator determined by different methods (activities are presented in mBq/kg, while the surface contamination by $^{207}$Bi is given in mBq/cm$^2$). Data for
$^{116}$CdWO$_4$ and CdWO$_4$ crystal scintillators are presented
for comparison.}
\begin{center}
\begin{tabular}{|l|l|l|l|l|}
 \hline
  Chain       & Nuclide     & \multicolumn{3}{c|}{Activity} \\
 \cline{3-5}
 ~          & ~           & $^{106}$CdWO$_4$    & $^{116}$CdWO$_4$              & CdWO$_4$ \\
 ~          & ~           & ~                   & \cite{Dane03a,Dane03b,Barab11} & \cite{Dane96a,Bell07a} \\
 \hline
 $^{232}$Th & $^{232}$Th  & $\leq0.07^{~a}$     & $\leq0.08-0.053(9)$           & $\leq0.026$ \\
 ~          & $^{228}$Th  & 0.042(4)$^{~b}$     & $0.039(2)-0.062(6)$           & $\leq(0.003-0.014)$ \\
 \hline
 $^{238}$U  & $^{238}$U   & $\leq 0.6^{~a}$     & $\leq (0.4-0.6)$              & $\leq 1.3$ \\
 ~          & $^{230}$Th  & $\leq0.4^{~a}$      & $\leq (0.05-0.5)$             & ~ \\
 ~          & $^{226}$Ra  & $0.012(3)^{~c}$     & $\leq 0.005$                  & $\leq(0.007-0.018)$ \\
 ~          & $^{210}$Po  & $\leq 0.2^{~a}$     & $\leq (0.063-0.6)$            & $\leq 0.063$ \\
 \hline
 Total $\alpha$ activity & ~ & 2.1(2)$^{~a}$    & $1.4(1)-2.7(3)$               & 0.26(4) \\
 \hline
 ~          & $^{40}$K    & $\leq 1.4^{~d}$       & $\leq 0.4$                    & $\leq(1.7-5)$ \\
 ~          & $^{90}$Sr-$^{90}$Y & $\leq0.3^{~d}$ & $\leq 0.2$                  & $\leq 1$ \\
 ~          & $^{106}$Ru  & $\leq 0.02^{~d}$    & --                            & -- \\
 ~          & $^{110m}$Ag & $\leq 0.06^{~d}$    & 0.06(4)                       & -- \\
 ~          & $^{113}$Cd  & 182$^{~e}$          & 91(5)                         & $558(4)-580(20)$ \\
 ~          & $^{113}$Cd$^m$ & 116000(4000)$^{~d}$ & 0.43(6)                       & $\leq 3.4-150(10)$ \\
 ~          & $^{137}$Cs  & --                  & 2.1(5)                        & $\leq 0.3$ \\
 ~          & $^{207}$Bi internal & $\leq 0.7^{~d}$     & 0.6(2)                        & -- \\
 ~          & $^{207}$Bi surface & $0.06^{~d}$     & --                        & -- \\

\hline
 \multicolumn{5}{l}{$^{a)}$ Pulse-shape discrimination (Section \ref{PSD})} \\
 \multicolumn{5}{l}{$^{b)}$ Time-amplitude analysis (Section  \ref{t-A})} \\
 \multicolumn{5}{l}{$^{c)}$ Analysis of double pulses (Section  \ref{bi-po})} \\
 \multicolumn{5}{l}{$^{d)}$ Fit of the background spectrum (Section  \ref{bg-sim})}\\
 \multicolumn{5}{l}{$^{e)}$ Calculated taking into account the isotopic abundance
 of $^{113}$Cd in $^{106}$CdWO$_4$ \cite{Bell10a}} \\
 \multicolumn{5}{l}{and the half-life of $^{113}$Cd \cite{Bell07a}.} \\
\end{tabular}
 \label{rad-cont}
\end{center}
\end{table}

\section{RESULTS AND DISCUSSION}

There are no peculiarities in the data accumulated with the
$^{106}$CdWO$_4$ detector which could be ascribed to the double
$\beta$ decay of $^{106}$Cd. Therefore only lower half-life limits
can be set by using the formula:

\begin{center}
$$\lim T_{1/2} = N \times \eta \times t \times \ln 2 / \lim S,$$
\end{center}

\noindent where $N$ is the number of $^{106}$Cd nuclei in the
$^{106}$CdWO$_4$ crystal ($2.42\times10^{23}$), $\eta$ is the
detection efficiency, $t$ is the time of measurements, and $\lim
S$ is the number of events of the effect searched for, which can be
excluded at a given confidence level (C.L.; all the limits on the
double beta processes in $^{106}$Cd are given
at 90\% C.L. in the present study).

\begin{figure}[htb]
\begin{center}
\mbox{\epsfig{figure=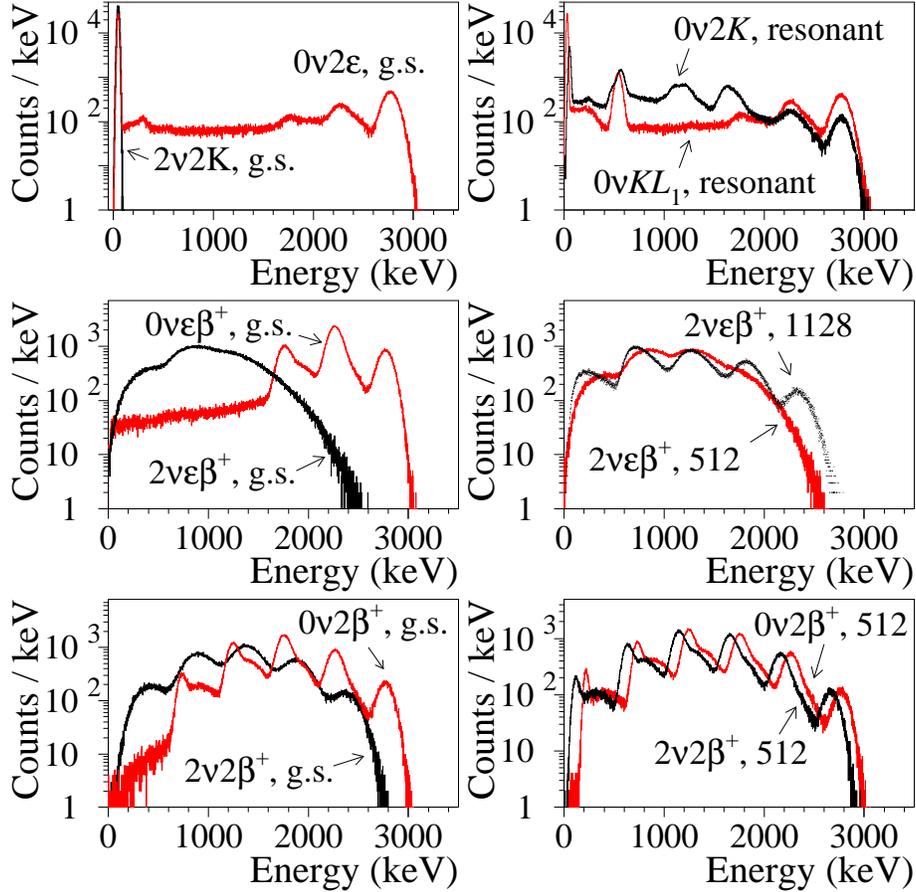,height=12.0cm}} \caption{(Color
online) Simulated response functions of the $^{106}$CdWO$_4$
detector to $2\varepsilon$, $\varepsilon\beta^+$ and $2\beta^+$
processes in $^{106}$Cd.}
 \label{fig:sim}
\end{center}
\end{figure}

The response functions of the $^{106}$CdWO$_4$ detector to the
2$\beta$ processes in $^{106}$Cd were simulated with the help of
the EGS4 \cite{EGS4} and the DECAY0 \cite{DECAY4} packages (some
examples of the simulated spectra are presented in Fig. \ref{fig:sim}).

\subsection{Double beta processes in $^{106}$Cd with positron(s) emission}

To estimate the value of $\lim S$ for the $2\nu\varepsilon\beta^+$
decay of $^{106}$Cd to the ground state of $^{106}$Pd, the energy
spectrum of the $\gamma$ and $\beta$ events accumulated over 6590 h
with the $^{106}$CdWO$_4$ detector was fitted by the model
built from the components of the background (see Section
3.4) and the effect searched for. The activities of U/Th daughters
in the crystals were constrained in the fit taking into account the
results of the time-amplitude and pulse-shape analyses.
The initial values of the $^{40}$K, $^{232}$Th and $^{238}$U activities
inside the PMTs were taken from \cite{Bern99}, where the radioactive
contaminations of PMTs of the same model were measured. The radioactive
contaminations of the copper were constrained taking into account the data
of the measurements \cite{Klap97} where copper of a similar quality was used.
The best fit (achieved in the
energy interval $780-2800$ keV with $\chi^2/$n.d.f.~$=93/81=1.15$)
gives an area of the $2\nu\varepsilon\beta^+$ distribution in the interval of the fit:
$(26\pm230)$ counts, thus with no evidence for the effect. In
accordance with the Feldman-Cousins procedure \cite{Feld98}, this
corresponds to $\lim S=403$ counts at 90\% C.L. Taking into
account the detection efficiency within the fit window given by the Monte
Carlo simulation ($\eta=0.700$) and the 98\% efficiency of the pulse-shape
discrimination to select $\gamma$($\beta$) events, we get the
following limit on the decay:

\begin{center}
$T_{1/2}^{2\nu\varepsilon\beta^+}($g.s.~$ \rightarrow $~g.s.$)
\geq 2.1\times10^{20}$ yr~~~~~at 90\% C.L.
\end{center}

\begin{figure}[htb]
\begin{center}
\mbox{\epsfig{figure=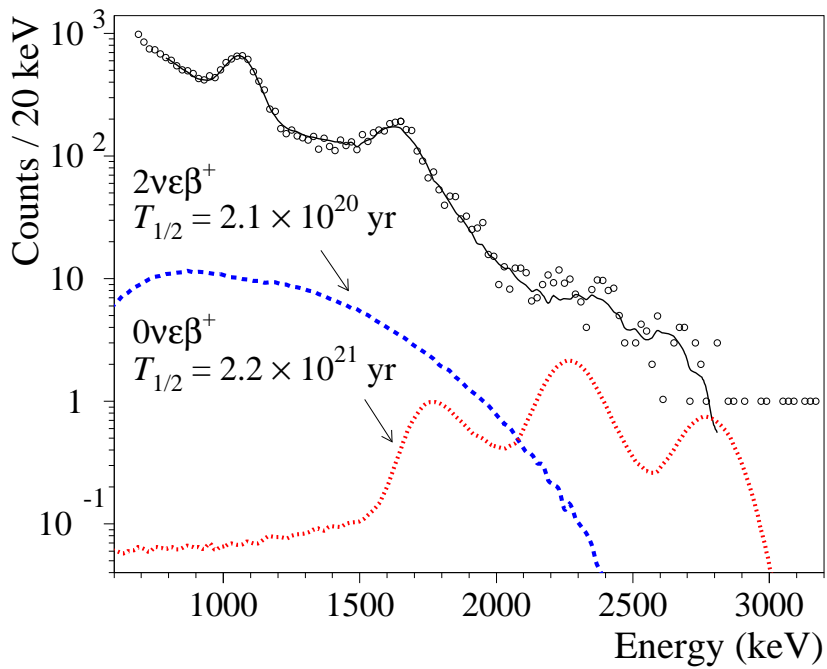,height=9.0cm}} \caption{(Color
online) Part of the energy spectrum of $\gamma$ and $\beta$ events
accumulated with the $^{106}$CdWO$_4$ detector over 6590 h
(circles) and its fit in the energy interval $780-2800$ keV (solid
line) together with the excluded distributions of $2\nu
\varepsilon\beta^+$ and $0\nu \varepsilon\beta^+$ decay of
$^{106}$Cd.}
 \label{fig:fit-2b}
\end{center}
\end{figure}

The excluded energy distribution expected for the two neutrino
$\varepsilon\beta^{+}$ decay of $^{106}$Cd is shown in Fig.
\ref{fig:fit-2b}.

One can prove this result by using the so called ``one sigma''
approach when a value of $\lim S$ can be estimated as the square root
of the counts in the energy interval of interest. There are 5462 events
in the energy interval $1140-2220$ keV where the detection
efficiency for the $2\nu\varepsilon\beta^+$ decay is 36\%. The
method gives a limit $T_{1/2}^{2\nu\varepsilon\beta^+}\geq
6.0\times10^{20}$ yr at 68\% C.L., similar to the result
acquired by fitting the experimental data with the help of the
Monte Carlo simulated models.

The sensitivity to the neutrinoless channel of the $\varepsilon\beta^+$
decay is better thanks to the shift of the energy distribution to
higher energies. Moreover, there are clear peaks in the spectrum
of the $0\nu\varepsilon\beta^+$ process in the energy region
$1.6-2.9$ MeV, which make the effect much more distinguishable
(see Fig. \ref{fig:fit-2b}). A fit of the data
in the energy interval $2000-3000$ keV
($\chi^2/$n.d.f.~$=23/25=0.92$) gives an area of the effect $(17\pm
13)$ events ($\lim S=38$ events, $\eta=0.675$), which corresponds to the
following limit on the $0\nu\varepsilon\beta^+$ decay of
$^{106}$Cd to the ground state of $^{106}$Pd:

\begin{center}
$T_{1/2}^{0\nu\varepsilon\beta^+}($g.s.~$ \rightarrow $~g.s.$)
\geq 2.2\times10^{21}$ yr~~~~~at 90\% C.L.
\end{center}

The ``one sigma''
approach gives for this decay the limit
(there are 187 events
in the energy interval $2140-2960$ keV, where the detection
efficiency for the $0\nu\varepsilon\beta^+$ decay is 62\%):
$T_{1/2}^{0\nu\varepsilon\beta^+}\geq
5.6\times10^{21}$ yr at 68\% C.L., proving the result
obtained by fitting the experimental data.

A fit of the data in the energy interval $1200-3000$ keV
($\chi^2/$n.d.f.~$=71/65=1.09$) gives $S=(92\pm52)$ events ($\lim
S=177$ events, $\eta=0.616$) of the $2\nu2\beta^+$ decay of $^{106}$Cd to the
ground level of $^{106}$Pd. Therefore, we set the following limit:

\begin{center}
$T_{1/2}^{2\nu2\beta^+}($g.s.~$ \rightarrow $~g.s.$) \geq
4.3\times10^{20}$ yr~~~~~at 90\% C.L.
\end{center}

The neutrinoless double positron decay was restricted by the fit
in the energy interval $760-2800$ keV
($\chi^2/$n.d.f.~$=88/82=1.07$, $S=-14\pm69$, $\lim S=100$, $\eta=0.956$):

\begin{center}
$T_{1/2}^{0\nu2\beta^+}($g.s.~$ \rightarrow $~g.s.$) \geq
1.2\times10^{21}$ yr~~~~~at 90\% C.L.
\end{center}

It should be stressed that the mass difference between $^{106}$Cd
and $^{106}$Pd atoms also allows transitions to the excited levels of
$^{106}$Pd. Thus, we have given limits on the $2\beta^+$ decay of
$^{106}$Cd to the first excited level of $^{106}$Pd ($2^+$, 512
keV), and on the electron capture with positron emission to a few
lowest excited levels of $^{106}$Pd with the spin-parity $0^+$ and
$2^+$. The results are presented in Table \ref{2b-results}.

\subsection{Double electron capture in $^{106}$Cd}

In the case of 2$\nu$ double electron capture in $^{106}$Cd from
the $K$ or/and $L$ shells the total energy release in the
$^{106}$CdWO$_4$ detector is in the range from $2E_{L3}=6.3$ keV
to $2E_K=48.8$ keV (where $E_K$ and $E_L$ are the binding energies
of the electrons on the $K$ and $L$ shells of the palladium atom,
respectively). Detection of such an energy deposit requires a low
enough energy threshold and low background conditions. In our
measurements the energy threshold for the acquisition was set too high
(because of the background due to the $\beta$ decay of $^{113}$Cd$^m$)
to search for the two neutrino mode of double electron capture to
the ground state and to the first excited level of $^{106}$Pd.

However, we can analyze the existing data to search for the $2\nu$
double electron capture to the higher excited levels of $^{106}$Pd.
For instance, by fitting the background spectrum in the energy
interval $660-2780$ keV ($\chi^2/$n.d.f.~$=103/86=1.20$, $S=-5\pm63$, $\lim S=99$, $\eta=0.328$) the following half-life limit on
$2\nu2\varepsilon$ decay of $^{106}$Cd to the 2$^{+}_{2}$ level
(1128 keV) of $^{106}$Pd was obtained:

\begin{center}
$T_{1/2}^{2\nu2\varepsilon}($g.s.~$ \rightarrow 2^{+}_{2})\geq
4.1\times10^{20}$ yr~~~~~at 90\% C.L.
\end{center}

The following restriction was set on the $2\nu2\varepsilon$ decay
of $^{106}$Cd to the $0^{+}_{1}$ 1134 keV level of $^{106}$Pd by
fitting the experimental spectrum in the energy interval 660--2800
($\chi^2/$n.d.f.~$=105/87=1.21$, $S=-5\pm163$, $\lim S=263$, $\eta=0.367$):

\begin{center}
$T_{1/2}^{2\nu2\varepsilon}(g.s. \rightarrow 0^{+}_{1})\geq
1.7\times10^{20}$ yr~~~~~at 90\% C.L.
\end{center}

In the case of the neutrinoless double electron capture, different
particles can be emitted: X rays and Auger electrons from
de-excitations in atomic shells, $\gamma$ quanta and/or conversion
electrons from de-excitation of daughter nucleus. We suppose here
that only one $\gamma$ quantum is emitted in the nuclear
de-excitation process. It should be stressed that the electron
captures from different shells ($2K$, $KL$, $2L$ and other modes)
cannot be energetically resolved by our detector. The fit of the
measured spectrum in the energy interval $1800-3200$ keV ($\chi^2/$n.d.f.~$=37/41=0.90$, $S=7\pm10$, $\lim S=23$, $\eta=0.194$) gives the
following limit on the $0\nu2\varepsilon$ transition of $^{106}$Cd to
the ground state of $^{106}$Pd:

\begin{center}
$T_{1/2}^{0\nu2\varepsilon}($g.s.~$ \rightarrow $~g.s.$)\geq
1.0\times10^{21}$ yr~~~~~at 90\% C.L.
\end{center}

The limits on the double electron capture in $^{106}$Cd to the lowest
excited levels of $^{106}$Pd were obtained by a fit of the data in
different energy intervals (see Table \ref{2b-results}).

\subsection{Resonant neutrinoless double electron capture in $^{106}$Cd}

A resonant
neutrinoless double electron capture in $^{106}$Cd is possible on
three excited levels of $^{106}$Pd with energies 2718 keV, 2741 keV and 2748
keV.

\begin{figure}[htb]
\begin{center}
\mbox{\epsfig{figure=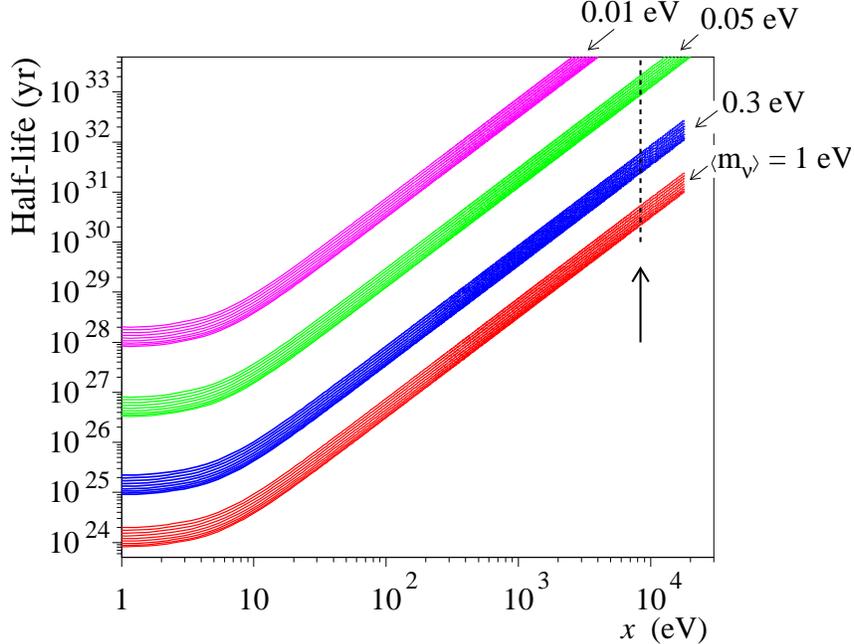,height=9.0cm}} \caption{(Color online) Calculated
half-life for the resonant $0\nu2\varepsilon$
capture decay of $^{106}$Cd to the excited level 2718 keV of
$^{106}$Pd as a function of parameter $x$ (see text) for different
values of the effective neutrino mass. Dashed line and arrow show the value of $x$ derived from the recent measurements of the $Q_{2\beta}$ in $^{106}$Cd \cite{Gon2011}.}
 \label{fig:theory}
\end{center}
\end{figure}

The half-life of the $^{106}$Cd resonant $2\varepsilon$ process
was estimated \cite{Suho11} by using the general formalism of
\cite{Suho98} and by calculating the associated nuclear matrix
element in a realistic single-particle space with a microscopic
nucleon-nucleon interaction. We have used a higher-RPA
(random-phase approximation) framework called the
multiple-commutator model (MCM) \cite{Suho93,Civi94}. Using the
UCOM short-range correlations \cite{Kort07}, the half-life for the
$0\nu$ double electron capture in $^{106}$Cd to the 2718 keV level
of $^{106}$Pd (assuming its spin-parity is $0^+$) can be written
as:

\begin{center}
$T_{1/2} = (3.0-8.1) \times 10^{22} \times \frac{x^2 +
26.2}{{\left\langle m_{\nu} \right\rangle}^2}$ yr ~~~~~~~~~~(1)
\end{center}

\noindent where $x = \left|Q_{2\beta}-E\right|$, and $\left\langle
m_{\nu} \right\rangle$ (the effective Majorana neutrino mass) are
in eV units. Here $Q_{2\beta}$ is the difference in atomic
masses between $^{106}$Cd and $^{106}$Pd, and $E$ contains the nuclear
excitation energy and the hole energies in the atomic $s$
orbitals. The dependence of the half-life on $x$
is plotted in Fig. \ref{fig:theory} for
several values of $\left\langle m_{\nu} \right\rangle$.
Use of the the recently remeasured (by the Penning-trap mass spectrometry \cite{Gon2011})
value of $Q_{2\beta}$ leads to a value $x=8390$ eV for the degeneracy parameter, and
thus to the 2$\varepsilon$ half-life estimate:
$T_{1/2} = (2.1 - 5.7) \times 10^{30}$ yr for $\left\langle m_{\nu}
\right\rangle=1$ eV.

We have estimated limits on the resonant 0$\nu2K$ and 0$\nu KL$
processes in $^{106}$Cd by using the data from our experiment. For
instance, the fit of the energy spectrum of the $\gamma$ and $\beta$
events measured by the $^{106}$CdWO$_4$ detector over 6590 h in
the energy region $1280-3000$ keV ($\eta=0.315$) gives $35\pm34$ events for the
$0\nu$ double electron captures from two $K$ shells to the excited
level at 2718 keV. We should take $\lim S=91$ events, which leads to
the following limit on the possible resonant process:

\begin{center}
$T_{1/2}^{0\nu 2K}($g.s.~$ \rightarrow $~2718 keV$)\geq
4.3\times10^{20}$ yr~~~~~at 90\% C.L.
\end{center}

\noindent For the $0\nu$ double electron capture of $K$ and $L_{1}$
electrons to the level 2741 keV we have obtained a slightly stronger
restriction ($S=10\pm13$, $\lim S=31$, $\eta=0.238$):

\begin{center}
$T_{1/2}^{0\nu KL_{1}}($g.s.~$ \rightarrow $~2741 keV$)\geq
9.5\times10^{20}$ yr~~~~~at 90\% C.L.
\end{center}

\noindent However, one can expect that the $0\nu KL$ process is
strongly suppressed due to the large spin ($4^+$) of the level at
2741 keV.

Finally, for the $0\nu$ double electron capture of $K$ and $L_{3}$
electrons to the $2,3^{-}$ level at 2748 keV we have obtained the following limit ($S=35\pm21$, $\lim S=69$, $\eta=0.238$):

\begin{center}
$T_{1/2}^{0\nu KL_{3}}($g.s.~$ \rightarrow $~2748 keV$)\geq
4.3\times10^{20}$ yr~~~~~at 90\% C.L.
\end{center}

Despite the fact that the limits are far away from the theoretical predictions, they
are higher than the existing limits and are at the level of the
best restrictions on resonant processes reported for different
isotopes. The limit for the $0\nu$ double electron capture to the level at 2748 keV is obtained for the first time.

All the half-life limits on $2\beta$ decay of $^{106}$Cd obtained
in the present work are summarized in Table \ref{2b-results} where
results of the most sensitive previous studies are given for
comparison.

\begin{table}[h!]
\caption{Half-life limits on 2$\beta$ processes in $^{106}$Cd. The detection
efficiencies for the effect searched for ($\eta$) and the values of $\lim S$
within the energy intervals of fit ($\Delta E$) are presented.}
\begin{center}
\begin{tabular}{|l|l|l|l|l|l|l|l|}
\hline
 Decay                  & Decay     & Level             & $\Delta E$ (keV)  & $\eta$    & $\lim S$  & \multicolumn{2}{c|}{ $T_{1/2}$ limit (yr) at 90\% C.L.}\\
\cline{7-8}
 channel                & mode      & of $^{106}$Pd     & ~             & ~         & ~         & Present work              & Best previous \\
 ~                      & ~         & (keV)             & ~             & ~         & ~         & ~                         & limits \\
 \hline
 $2\varepsilon$         & $2\nu $   & g.s.              & $-$           & $-$       & $-$       & $-$                       & $\geq3.6\times10^{20}$  \cite{Rukh11a} \\
 ~                      & ~         & 2$^{+}_{1}$ 512   & $-$           & $-$       & $-$       & $-$                       & $\geq1.2\times10^{20}$  \cite{Rukh11b} \\
 ~                      & ~         & 2$^{+}_{2}$ 1128  & $660-2780$        & 0.328 & 99        & $\geq 4.1\times10^{20}$   & $\geq5.1\times10^{18}$  \cite{Bara96a} \\
 ~                      & ~         & 0$^{+}_{1}$ 1134  & $660-2800$        & 0.367 & 263       & $\geq 1.7\times10^{20}$   & $\geq1.0\times10^{20}$  \cite{Rukh11b} \\
 ~                      & ~         & 2$^{+}_{3}$ 1562  & $660-2800$        & 0.342 & 830       & $\geq 5.1\times10^{19}$   & $-$ \\
 ~                      & ~         & 0$^{+}_{2}$ 1706  & $760-2800$        & 0.320 & 370       & $\geq 1.1\times10^{20}$   & $-$ \\
 ~                      & ~         & 0$^{+}_{3}$ 2001  & $760-2780$        & 0.484 & 208       & $\geq 2.9\times10^{20}$   & $-$ \\
 ~                      & ~         & 0$^{+}_{4}$ 2278  & $660-3000$        & 0.381 & 294       & $\geq 1.6\times10^{20}$   & $-$ \\
 \cline{2-8}
 ~                      & $0\nu$    & g.s.              & $1800-3200$       & 0.194 & 23        & $\geq 1.0\times10^{21}$   & $\geq8.0\times10^{18}$  \cite{Dane03b} \\
 ~                      & ~         & 2$^{+}_{1}$ 512   & $2040-3200$       & 0.150 & 36        & $\geq 5.1\times10^{20}$   & $\geq3.5\times10^{18}$  \cite{Bara96a} \\
 ~                      & ~         & 2$^{+}_{2}$ 1128  & $760-3000$        & 0.465 & 187       & $\geq 3.1\times10^{20}$   & $\geq4.9\times10^{19}$  \cite{Bell99} \\
 ~                      & ~         & 0$^{+}_{1}$ 1134  & $760-3000$        & 0.474 & 169       & $\geq 3.5\times10^{20}$   & $\geq7.3\times10^{19}$  \cite{Bell99} \\
 ~                      & ~         & 2$^{+}_{3}$ 1562  & $760-3000$        & 0.520 & 186       & $\geq 3.5\times10^{20}$   & $-$ \\
 ~                      & ~         & 0$^{+}_{2}$ 1706  & $760-3000$        & 0.531 & 262       & $\geq 2.5\times10^{20}$   & $-$ \\
 ~                      & ~         & 0$^{+}_{3}$ 2001  & $1300-3200$       & 0.346 & 185       & $\geq 2.3\times10^{20}$   & $-$ \\
 ~                      & ~         & 0$^{+}_{4}$ 2278  & $660-3200$        & 0.564 & 335       & $\geq 2.1\times10^{20}$   & $-$ \\
 \hline
 Res. $2K$              & $0\nu$    & ~~~~~~~2718       & $1280-3000$       & 0.315 & 91        & $\geq 4.3\times10^{20}$   & $\geq1.6\times10^{20}$  \cite{Rukh11b} \\
 Res. $KL_{1}$          & ~     & $4^+$ ~~~2741     & $1280-3000$       & 0.238 & 31        & $\geq 9.5\times10^{20}$   & $\geq1.1\times10^{20}$  \cite{Rukh11a} \\
 Res. $KL_{3}$          & ~     & $2,3^-$ 2748      & $1300-3000$       & 0.238 & 69        & $\geq 4.3\times10^{20}$   & $-$ \\
 \hline
 $\varepsilon\beta^+$   & $2\nu$    & g.s.              & $780-2800$        & 0.700 & 403       & $\geq 2.1\times10^{20}$   & $\geq4.1\times10^{20}$  \cite{Bell99} \\
 ~                      & ~         & 2$^{+}_{1}$ 512   & $660-3000$        & 0.846 & 943       & $\geq 1.1\times10^{20}$   & $\geq2.6\times10^{20}$  \cite{Bell99} \\
 ~                      & ~         & 2$^{+}_{2}$ 1128  & $1260-3000$       & 0.414 & 167       & $\geq 3.1\times10^{20}$   & $\geq1.4\times10^{20}$  \cite{Bell99} \\
 ~                      & ~         & 0$^{+}_{1}$ 1134  & $1200-3000$       & 0.519 & 172       & $\geq 3.7\times10^{20}$   & $\geq1.6\times10^{20}$  \cite{Rukh11b}\\
 \cline{2-8}
 ~                      & $0\nu$    & g.s.              & $2000-3000$       & 0.675 & 38        & $\geq 2.2\times10^{21}$   & $\geq3.7\times10^{20}$  \cite{Bell99} \\
 ~                      & ~         & 2$^{+}_{1}$ 512   & $1200-3000$       & 0.936 & 91        & $\geq 1.3\times10^{21}$   & $\geq2.6\times10^{20}$  \cite{Bell99} \\
 ~                      & ~         & 2$^{+}_{2}$ 1128  & $1200-3000$       & 0.678 & 148       & $\geq 5.7\times10^{20}$   & $\geq1.4\times10^{20}$  \cite{Bell99} \\
 ~                      & ~         & 0$^{+}_{1}$ 1134  & $2000-3000$       & 0.240 & 59        & $\geq 5.0\times10^{20}$   & $\geq1.6\times10^{20}$  \cite{Rukh11b} \\
 \hline
 $2\beta^+$             & $2\nu$    & g.s.              & $1200-3000$       & 0.616 & 177       & $\geq 4.3\times10^{20}$   & $\geq2.4\times10^{20}$  \cite{Bell99} \\
  ~                     & ~         & 2$^{+}_{1}$ 512   & $760-2800$        & 0.831 & 203       & $\geq 5.1\times10^{20}$   & $\geq1.7\times10^{20}$  \cite{Rukh11b} \\
 \cline{2-8}
 ~                      & $0\nu$    & g.s.              & $760-2800$        & 0.956 & 100       & $\geq 1.2\times10^{21}$   & $\geq2.4\times10^{20}$  \cite{Bell99} \\
 ~                      & ~         & 2$^{+}_{1}$ 512   & $780-3000$        & 0.870 & 92        & $\geq 1.2\times10^{21}$   & $\geq1.7\times10^{20}$  \cite{Rukh11b} \\
 \hline
\end{tabular}
\label{2b-results}
\end{center}
\end{table}

Although the obtained bounds are well below the existing
theoretical predictions
\cite{Hirs94,Stau91,Toiv97,Stoi03,Shuk05,Domi05}, most of the
limits are about one order of magnitude higher than those previously
established. Moreover, some channels of $^{106}$Cd
double $\beta$ decay were investigated for the first time. It
should be stressed that only two nuclides ($^{78}$Kr \cite{Gavr11}
and $^{130}$Ba \cite{Mesh01}) among six potentially $2\beta^+$
active isotopes \cite{DBD-tab} were investigated at a comparable
level of sensitivity $T_{1/2}\sim$$10^{21}$ yr.

A new phase of the experiment with the $^{106}$CdWO$_4$
scintillation detector placed in the ultra-low background GeMulti
set-up (four HPGe detectors of 225 cm$^3$ volume each,
located at the Gran Sasso National Laboratories) is in
preparation. We are going to record pulse-profiles and arrival
time of the events from the $^{106}$CdWO$_4$ scintillator both in
coincidence and anti-coincidence modes. To suppress the background due
to the radioactive contamination of the PMT, the development of a lead
tungstate (PbWO$_4$) active light-guide from ultra-pure
archaeological lead \cite{Dane09,Kovt11} has been completed. Our
preliminary simulations show that such an experiment could
investigate the $2\nu$ mode of $\varepsilon\beta^+$ and of $2\beta^+$
decays, and also $2\varepsilon$ transitions of $^{106}$Cd to
the excited states of $^{106}$Pd, at a level of sensitivity near to
the theoretical predictions: $T_{1/2} \sim 10^{20}-10^{22}$ yr
\cite{Hirs94,Stau91,Toiv97,Stoi03,Shuk05,Domi05}.

Moreover, the development of a $^{106}$CdWO$_4$ crystal
scintillator depleted in the $^{113}$Cd isotope by a factor
$10^{3}-10^{4}$ (to reduce the background caused by $\beta$ decay
of $^{113}$Cd$^m$) is also possible \cite{Tikhomirov}. Such a
detector could be able to investigate two neutrino double electron
capture, which is theoretically the most favorable process of
$2\beta$ decay of $^{106}$Cd.

\section{CONCLUSIONS}

A low background experiment using radiopure cadmium tungstate
crystal scintillator (215 g) enriched in $^{106}$Cd to 66$\%$ has
been carried out at the underground Gran Sasso National Laboratories
of the INFN. The background of the detector below 0.65 MeV is
mainly due to the $\beta$ active $^{113}$Cd$^m$ ($\approx 116$ Bq/kg). We
have found surface contamination of the crystal by $^{207}$Bi at
level of 3 mBq, which provides a considerable part of the
background up to $\approx2.5$ MeV. The activities of U/Th in the
scintillator are rather low: $\approx0.04$ mBq/kg of $^{228}$Th
and $\approx0.01$ mBq/kg of $^{226}$Ra. The total $\alpha$ activity of
U/Th is at level of $\approx 2$ mBq/kg. A background counting
rate of the detector in the vicinity of the $^{106}$Cd double beta
decay energy ($2.7-2.9$ MeV), after rejection of
$^{212}$Bi~--~$^{212}$Po events, is 0.4
counts/(yr$\times$keV$\times$kg).

After 6590 h of data taking, new improved limits on $2\beta$ decay
of $^{106}$Cd were set at level of $10^{19}-10^{21}$ yr, in
particular: $T_{1/2}^{2\nu\varepsilon\beta^+}\geq2.1\times
10^{20}$~yr, $T_{1/2}^{2\nu2\beta^+}\geq4.3\times 10^{20}$~yr, and
$T_{1/2}^{0\nu2\varepsilon}\geq1.0\times 10^{21}$~yr. Resonant
$0\nu2\varepsilon$ processes have been restricted to:
$T_{1/2}^{0\nu2K}($g.s.~$ \rightarrow $~2718 keV$)\geq4.3\times 10^{20}$~yr; $T_{1/2}^{0\nu
KL_{1}}($g.s.~$ \rightarrow $~2741 keV$)\geq9.5\times 10^{20}$~yr and $T_{1/2}^{0\nu
KL_{3}}($g.s.~$ \rightarrow $~2748 keV$)\geq4.3\times 10^{20}$~yr (all the limits at 90\%~C.L.). A
possible resonant enhancement of $0\nu2\varepsilon$ processes was
estimated in the framework of the QRPA approach. The half-life of the
resonant decay depends on the difference between the value of
$Q_{2\beta}$ and of the energies of the appropriate excited levels of
$^{106}$Pd minus the binding energies of two electrons on shells of
the daughter atom. The half-life decreases with the decrease of this difference.

A next stage of the experiment is in preparation. We are going to
install a low background scintillation detector with the
$^{106}$CdWO$_4$ crystal into the GeMulti ultra-low background
set-up with four 225 cm$^3$ HPGe detectors at the Gran Sasso
National Laboratories. The sensitivity of the experiment, in
particular to the two neutrino $\varepsilon\beta^+$ decay of
$^{106}$Cd, is expected to be enhanced thanks to the high energy
resolution of the GeMulti detector and to the improvement of the background
conditions in coincidence mode. In addition, we hope to reduce the surface
contamination of the scintillator with $^{207}$Bi,
observed in the present study, by cleaning (removing) the crystal surface.
We estimate the sensitivity of the experiment, in
particular to the $2\nu\varepsilon\beta^+$ decay of $^{106}$Cd, to
be at level of the theoretical predictions $T_{1/2} \sim
10^{20}-10^{22}$ yr.

Moreover, a further improvement of sensitivity can be reached by
increasing the enrichment factor of $^{106}$Cd, and by developing
$^{106}$CdWO$_4$ scintillators with lower level of radioactive
contaminations, including depletion in $^{113}$Cd. A
$^{106}$CdWO$_4$ scintillation detector with
an activity of $^{113}$Cd$^m$ reduced by a factor
of $10^{3}-10^{4}$ could be able to detect
two neutrino double electron capture in $^{106}$Cd, which is
theoretically the most probable process.

\section{ACKNOWLEDGMENTS}

The group from the Institute for Nuclear Research (Kyiv, Ukraine)
was supported in part by the Project "Kosmomikrofizyka-2"
(Astroparticle Physics) of the National Academy of Sciences of
Ukraine. D.V.~Poda and O.G.~Polischuk were supported in part by
the Project "Double beta decay and neutrino properties" for young
scientists of the National Academy of Sciences of Ukraine (Reg.
No. 0110U004150). Authors would like to express their gratitude to the
Referee for the careful reading of the manuscript and for
the valuable comments.

\end{document}